\documentclass[prd,12pt,showpacs,aps,nofootinbib]{revtex4-1}
\usepackage{amssymb}
\usepackage{amsmath,amssymb,graphicx}
\usepackage{graphicx}
\usepackage{dcolumn}
\usepackage{bm}
\def\be{\begin{equation}}
\def\ee{\end{equation}}
\def\bea{\begin{eqnarray}}
\def\eea{\end{eqnarray}}

\usepackage{pstricks}
\usepackage{graphicx}
\usepackage{amsbsy}
\usepackage{bm}
\usepackage{color}
\usepackage{fancyhdr}
\usepackage{epsfig}
\usepackage{hyperref}

\begin{document}

\bigskip

\vspace{2cm}
\title{Relation between masses of particles and the Fermi constant in the electroweak Standard Model }
\vskip 6ex
\author{G. L\'opez Castro}
\email{glopez@fis.cinvestav.mx}
\affiliation{Departamento de F\'isica, Centro de Investigaci\'on y de Estudios Avanzados, Apartado Postal 14-740, 07000 M\'exico D.F., M\'exico}
\author{J. Pestieau}
\email{jean.pestieau@gmail.com}
\affiliation{Institut de Recherche en Math\'ematique et en Physique,  Universit\'e catholique de Louvain, 2, Chemin du Cyclotron, Louvain-la-Neuve, Belgium}

\begin{abstract}
 An empirical formula relating the physical masses of elementary particles and the Fermi constant is proposed. Although no mechanism or theoretical model behind this formula is advocated, we seek for a possible physical interpretation. If not a simple numerical coincidence, this formula may motivate theorists to search for relations among the different sectors of the Electroweak Standard model. 
 
\end{abstract}

\maketitle

  The finding of successful empirical relations among the physical parameters of a system have provided 
a valuable guide in the formulation of physical theories. Well known examples are the roles of Planck's formula for the black body radiation or Balmer's formula for the spectrum of hydrogen in the formulation of the quantum theory. Even if they were considered as fortuitous at the beginning, their successfully properties served as a suitable constraint to be reproduced by a deeper theory aiming to explain the related phenomena. 

  In the present note we propose an empirical relation between the physical masses of fermions and bosons and the (precisely measured) fundamental Fermi constant $G_F$ \cite{fermi} in the Standard electroweak Model. Although at present we do not know any mechanism or theoretical model giving rise to this relation, we consider it to be  
very appealing and worth to receive attention by particle theorists. We also address some comments aiming to seek for a possible interpretation of this mass relation.

  Using the measured values \cite{atlas-cms,top,pdg}of the fermions and bosons of the Standard Model, it is easy to check  that the following relation is satisfied
\be
m_H^2+m_W^2+m_Z^2+\sum_{\rm f=q,l} m_f^2=v^2\ , \label{relation}
\ee
where $v$ is the vacuum expectation value of the scalar field H \cite{scalar} and $W$ and $Z$ are the weak gauge bosons. The sum in the left-hand side extends over all quarks and leptons. Using $m_H=(125.6 \pm 0.4)$ GeV \cite{atlas-cms,top,other} and $m_t=(173.5\pm 0.9)$ GeV \cite{top,pdg} and other masses from the PDG compilation \cite{pdg}, the left-hand side of Eq. (\ref{relation}) is in full agreement with the right-hand side because $v^2=1/(\sqrt{2}G_F)=(246.2\ {\rm GeV})^2$ according to \cite{pdg,pich,weinberg}. Observe that in the left-hand side of Eq. (\ref{relation}) there are almost equal contributions of fermions and bosons. Thus, the Fermi constant $G_F$ gets fixed from the sum of the masses of all elementary particles.

  Another look at the above sum rule can be endowed by noticing that, in the electroweak model, the couplings of all particles to the $H$ field are proportional to $v$ \cite{pich,weinberg}, namely of the form $m_i/v$. If one divides  Eq. (\ref{relation}) by $v^2$ it yields:
\be
2\lambda+\frac{g^2}{4}+\frac{g^2+g'^2}{4}+\sum_f \frac{y_f^2}{2} =1\ , \label{couplings}
\ee 
where $\lambda, g,\ g'$ and $y_f$ being, respectively, the effective and renormalized \cite{fleischer} scalar, gauge and Yukawa couplings. 

  On the other hand, in the electroweak model the potential for the scalar doublet $\Phi$ is given by:
\be
V(\Phi)=  \lambda (\Phi^{\dagger}\Phi-v^2/2)^2\ ,
\ee
with $m_H^2=2\lambda v^2$, the square of the mass of the physical scalar. Therefore, if Eq. (\ref{relation}) is admitted, the scalar potential becomes interrelated to other sectors of the electroweak Lagrangian. Indeed all couplings are included in Eq. (\ref{couplings}) through $v^2$.

   The fulfillment of Eq. (\ref{relation}) may be just a numerical coincidence. Instead, let us try to speculate about its possible physical interpretation. First, let us invoke an analogy with general relativity where gravity is seen as the curvature of space-time, which in turn is produced by the density of energy/mass. This interpretation is radically different from Newtonian interaction in terms of  force fields that are generated by the distribution of matter.  Similarly, in the electroweak model after 1967 \cite{weinberg}, the masses of particles are seen as generated from their interaction with a scalar field which has a nonzero vacuum expectation value, a mechanism that differs from the original Glashow's model \cite{glashow} where the masses of elementary particles would have to be introduced by hand. In turn, in  Eq. (\ref{relation}) $v$  can be viewed as the yield of the masses of fundamental particles, all of them lighter than $v$, which result from the quantum fluctuations in the electroweak vacuum. On the other hand, given that $v^2=1/(\sqrt{2}G_F)$,  Eq. (\ref{relation}) lends the alternative interpretation that the Fermi constant may be fixed from the interactions of particles with the electroweak vacuum.

\medskip

We thank Vincent Lema\^itre, Jean-Marc G\'erard, Thomas Hambye and Jean-Marie
Fr\`re for their useful comments


\end{document}